\documentclass[showpacs,showkeys,notitlepage,12pt]{revtex4}%
\usepackage{amssymb}
\usepackage{amsfonts}
\usepackage{amsmath}
\usepackage{latexsym}
\usepackage{graphicx}%
\begin{document}
\title{Bosonic (meta)stabilization of cosmic string loops\\ }
\author{J.R. Morris}
\affiliation{Physics Dept., Indiana University Northwest, 3400 Broadway, Gary, Indiana
46408 USA}

\begin{abstract}
We consider the possibility of a bosonic (meta)stabilization of a cosmic gauge
string loop due to the presence of a gas of low mass bosonic particles which
become trapped within the string core. This boson gas exerts a pressure which
tends to counteract the string tension, allowing a circular string loop to
find a finite equilibrium radius, provided that gas particles do not escape
the string core. However, high energy bosons do escape, and consequently the
loop shrinks and the temperature rises. Estimates indicate that the bosonic
stabilization mechanism is ineffective, and the loop is unstable against decay.

\end{abstract}

\pacs{11.30.Qc, 11.17.+y, 98.80.Cq}
\keywords{cosmic string loop, bosonic stabilization, domain kink}\maketitle

\section{Introduction}

\ \ The effects of fermions on the stability of nontopological defects, such
as semilocal and electroweak strings\cite{V-A PRD91},\cite{Vach PRL92}, has
been studied with some interesting results. For example, the presence of
fermions within the core of a nontopological string can have a stabilizing
effect\cite{Stojk1},\cite{SSV PRD01},\cite{SSV PRD02}. (See, e.g., \cite{A-C
rev} for an extensive review.) Presently, we investigate the possible
existence of a bosonic stabilizing effect for Abelian-Higgs string loops.

\bigskip

\ \ In Ref.\cite{Morris95} a model was introduced where, by the Witten
mechanism\cite{Witten}, a complex-valued gauge string scalar field $\chi$
interacts with a U(1) gauge field $A_{\mu}$ along with a \textit{real-valued}
scalar field $\phi$, instead of a complex scalar field. So, unlike the Witten
superconducting string model, there is only one U(1) gauge field, confined to
the string core, and the non-string scalar field $\phi$ is real rather than
complex. It was seen in \cite{Morris95} that for a certain range of parameters
a discrete $Z_{2}$ symmetry associated with the scalar $\phi$ could be
spontaneously broken within the string core, resulting in a set of vacuum
domains where $\phi=\pm\phi_{0}$. The $+\phi_{0}$ and $-\phi_{0}$ vacuua are
separated by topological solitonic \textquotedblleft kinks\textquotedblright%
\ and \textquotedblleft antikinks\textquotedblright\ described by $\phi_{K}$
and $\phi_{\bar{K}}$ interpolating between the vacuum states $\pm\phi_{0}$. If
the string is considered as a thin tube, the domain kinks can be thought of as
small domain wall sections within the string core, and in the limit that the
string is considered infinitely thin, the domain kinks are just ordinary kinks
of an effective 1+1 dimensional field theory in the string core.

\bigskip

\ \ The kinks (K) and antikinks (\={K}) will undergo annihilations and produce
elementary $\varphi$ bosons, which are the perturbative particle excitations
of the $\phi$ field. The $\varphi$ bosons can have a mass $m_{in}$ inside the
string core which is much less than the $\varphi$ boson mass $m=m_{out}$
outside the string, where the vacuum state is $\phi=0$. Thus, assuming this to
be the case, i.e., $m_{in}\ll m$, the $\varphi$ bosons will get effectively
trapped within the string, with a slow rate of leakage associated with the
escape of high energy $\varphi$ bosons with energies $E\gtrsim m$, if the
$\varphi$ gas temperature remains low, $\beta m\gg1$. (In Ref.\cite{Morris95}
the oppsosite case $m_{in}\gg m$ was assumed, resulting in $\varphi$ boson
radiation being released into the vacuum outside the string.) However, if
$m_{in}\ll m$, a mechanism can exist where the trapped $\varphi$ bosons create
what we may call here a \textquotedblleft\textit{pressurized}%
\textquotedblright\ string, for lack of a better name. Closed string loops can
form through intercommutations and self intersecting trajectories. The
$\varphi$ pressure in a string loop leads to a tendency for the loop to
expand, while the string tension tends to cause it to shrink. The result will
be a stabilization at some equilibrium radius $\mathcal{R}$ and length
$L=2\pi\mathcal{R}$ for a circular string loop. This string loop may actually
exist as a metastable state over longer periods of time, whose rate of decay
ultimately depends upon the rate of energy loss through gravitational and
$\chi$ boson radiation, as well as the rate of $\varphi$ particle leakage from
the string, which, in turn, depends upon $m_{in}/m$ and the $\varphi$ energy
distribution within the string.

\bigskip

\ \ However, an analysis of a loop's stability against decay due to $\varphi$
gas loss indicates that although the $\varphi$ gas may leak out at a
relatively slow rate at first, the rate rapidly increases due to an increasing
temperature. The lifetime of this bosonic stabilization mechanism, treating
the loop as a perfect $\phi$ blackbody for $\varphi$ particle energies $E\geq
m$, is found to be $\tau\sim m_{\chi}^{-1}$, where $m_{\chi}$ is the $\chi$
boson mass. This time $\tau$ is very short in comparison to the string's
lifetime due to gravitational radiation, and consequently the stabilization
mechanism is found to be ineffective, so that the loop remains unstable
against decay.

\bigskip

\ \ In section II a brief review of the field theory model describing kinks
nested within a host string is given, with K-\={K} annihilations contributing
to a confined gas of K-\={K} states and $\varphi$ bosons, producing a
pressurized string. For $m_{in}\ll M_{K}$, K-\={K} annihilations are much more
probable than K-\={K} pair productions from $\varphi$ bosons, and the gas
within the string core therefore consists mostly of $\varphi$ bosons. The
equilibrium conditions for a circular string loop are given in section III,
and the process of loop decay is studied in section IV. Particle masses and
model constraints are listed in section V. Section VI forms a brief summary.

\section{The model}

\subsection{Kinks nested within a gauge string}

\ \ \ \textbf{The gauge string -- }\ Let's first consider a straight Abelian
gauge string\cite{Nielsen,Vilenkin,VSbook} lying on the $z$ axis. There is a
complex-valued scalar string field $\chi=\frac{1}{\sqrt{2}}R\,\,e^{i\alpha}$
that interacts with the U(1) gauge field $A_{\mu}=\frac{1}{e}\left[  P(\vec
{x},t)-1\right]  \partial_{\mu}\alpha$ and also with a real-valued scalar
field $\phi$. The Lagrangian for this system is%

\begin{equation}
\label{one}L=(D^{\mu}\chi)^{\star}(D_{\mu}\chi)+\frac12\partial^{\mu}%
\phi\partial_{\mu}\phi-V-\frac14F^{\mu\nu}F_{\mu\nu}\text{,}%
\end{equation}

\noindent where the gauge covariant derivative $D_{\mu}$ and the U(1) field
tensor $F_{\mu\nu}$ are%

\begin{equation}
\label{two}D_{\mu}=\nabla_{\mu}+ieA_{\mu},\,\,\,\,\,\,\,\,\,\,F_{\mu\nu
}=\partial_{\mu}A_{\nu}-\partial_{\nu}A_{\mu}=\frac1e(\partial_{\mu}%
P\partial_{\nu}\alpha-\partial_{\nu}P\partial_{\mu}\alpha)\text{,}%
\end{equation}

\noindent The potential $V$ for the system takes the form%

\begin{equation}%
\begin{array}
[c]{c}%
V=\lambda\left(  \chi^{\star}\chi-\frac{1}{2}\eta^{2}\right)  ^{2}+f\left(
\chi^{\star}\chi-\frac{1}{2}\eta^{2}\right)  \phi^{2}+\frac{1}{2}m^{2}\phi
^{2}+\frac{1}{4}g\phi^{4}\\
=\frac{1}{4}\lambda\left(  R^{2}-\eta^{2}\right)  ^{2}+\frac{1}{2}f\left(
R^{2}-\eta^{2}\right)  \phi^{2}+\frac{1}{2}m^{2}\phi^{2}+\frac{1}{4}g\phi^{4}.
\end{array}
\label{three}%
\end{equation}

\noindent Here, $\nabla_{\mu}$ represents the ordinary spacetime covariant
derivative, and a metric with signature $(+,-,-,-)$ is used. The coupling
constants $\lambda$, $f$, and $g$, along with the $\phi$ particle mass $m$,
are taken to be positive, real quantities. The stable vacuum state of the
theory is located by $R=\eta$, $\phi=0$, and the $\chi$ particle mass is
$m_{\chi}=\left(  2\lambda\right)  ^{\frac{1}{2}}\eta$.

\bigskip

\ From (\ref{one}) the field equations for this system can be obtained:%

\begin{equation}
\label{four}\nabla_{\mu}\partial^{\mu}R-RP^{2}\partial_{\mu}\alpha
\partial^{\mu}\alpha+ \left[  \lambda\left(  R^{2}-\eta^{2}\right)  +f\phi
^{2}\right]  R=0,
\end{equation}

\begin{equation}
\label{five}\nabla_{\mu}\left(  \partial^{\mu}P\partial^{\nu}\alpha
-\partial^{\nu}P\partial^{\mu}\alpha\right)  +e^{2}R^{2}P\partial^{\nu}%
\alpha=0,
\end{equation}

\begin{equation}
\label{six}\nabla_{\mu}j^{\mu}=0,\,\,\,\,\,\,\,\,\,\,j_{\mu}=-R^{2}%
P\partial_{\mu}\alpha,
\end{equation}

\begin{equation}
\label{seven}\nabla_{\mu}\partial^{\mu}\phi+\left[  f\left(  R^{2}-\eta
^{2}\right)  +m^{2}+g\phi^{2}\right]  \phi=0.
\end{equation}

\noindent Now, when the real scalar field $\phi=0$, this model possesses a
solution set that describes an Abelian gauge string. With cylindrical
coordinates $(r,\theta,z)$ a static string lying on the $z$ axis is depicted
by the functions $R(r)$, $P(r)$, and $\alpha=\theta$. Theses are subject to
the boundary conditions $R(0)=0$, $P(0)=1$ in the center of the string core,
and asymptotically, outside the string core, $R\rightarrow\eta$,
$P\rightarrow0$. The radius of the string is $r_{0}\approx m_{\chi}%
^{-1}\approx(2\lambda\eta^{2})^{-\frac{1}{2}}$. For mathematical convenience
and simplicity, we consider the string as a tube of false vacuum of radius
$r_{0}$ with $R\approx0$ in the interior and $R\approx\eta$ outside.

\bigskip

However, when the scalar field $\phi$ does not vanish identically, there is a
parameter range for which it becomes energetically favorable for a $\phi$
condensate to form within the string core\cite{Morris95}. To see this, set
$R\approx0$ in the string core, and then the potential $V$ given by
(\ref{three}) is minimized by a field configuration for which $\phi=\pm
\phi_{0}$, where%

\begin{equation}
\label{eight}\phi_{0}=\left[  \left(  f\eta^{2}-m^{2}\right)  /g\right]
^{\frac12}%
\end{equation}

\noindent is a positive constant for $f\eta^{2}-m^{2}>0$. Then $V\left(
0,\pm\phi_{0}\right)  =\frac{1}{4}\left(  \lambda\eta^{4}-g\phi_{0}%
^{4}\right)  <V\left(  0,0\right)  =\frac{1}{4}\lambda\eta^{4}$. Of course the
field $\phi$ has gradient energy also, owing to the fact that $\phi
\rightarrow0$ outside the string. As was shown in \cite{Morris95} there is a
parameter range for which $\phi=0$ is unstable in the string core, and a
condensate with $\phi=\pm\phi_{0}$ forms there, breaking the $Z_{2}$ symmetry,
with $\phi\rightarrow0$ outside the string. We will assume that the model
parameters do indeed assume values that allow the formation of the $\phi
=\pm\phi_{0}$ condensate in the core.\bigskip

\ \ \textbf{Kinks --}\ \ \ \ We denote the static string background fields by
$R_{s}$ and $P_{s}$ and adopt the ansatz $R\approx R_{s}(r)$, $P\approx
P_{s}(r)$, $\alpha=\theta$, and $\phi=\phi(r,z,t)$. Then (\ref{seven})
becomes, approximately,%

\begin{equation}
\label{eleven}-\partial_{0}^{2}\phi+\partial_{r}^{2}\phi+\frac1r\partial
_{r}\phi+\partial_{z}^{2}\phi-\left[  g\left(  \phi^{2}-\phi_{0}^{2}\right)
+fR^{2}\right]  \phi=0,
\end{equation}

\noindent with the boundary conditions $\phi\rightarrow\phi(z,t)$ as
$r\rightarrow0$, $\phi\rightarrow0$ as $r\rightarrow\infty$, and $\left\vert
\phi\right\vert \rightarrow\phi_{0}$ as $\left\vert z\right\vert
\rightarrow\infty$. Using our tube representation for the string, we write%

\begin{equation}
\label{twelve}R\approx\left\{
\begin{array}
[c]{cc}%
0, & r\leq r_{0}\\
\eta, & r>r_{0}%
\end{array}
\right\}  ,\,\,\,\,\,\,\,\,\,\,\phi\approx\left\{
\begin{array}
[c]{cc}%
\phi(z,t), & r\leq r_{0}\\
0, & r>r_{0}%
\end{array}
\right\}
\end{equation}

\noindent so that in the string core (\ref{eleven}) simplifies to%

\begin{equation}
\label{thirteen}-\partial_{0}^{2}\phi+\partial_{z}^{2}\phi-g\phi\left(
\phi^{2}-\phi_{0}^{2}\right)  =0.
\end{equation}

\noindent A static solution of (\ref{thirteen}) is given by%

\begin{equation}
\label{fourteen}\phi_{K}(z)=\phi_{0}\tanh\left(  \frac{z-z_{0}}w\right)
,\,\,\,\,\,\,\,\,\,\,w=\frac1{\phi_{0}}\left(  \frac2g\right)  ^{\frac12},
\end{equation}

\noindent representing a static domain kink (K) of width $w$ located at the
position $z=z_{0}$ in the string. A domain antikink (\={K}) solution
\cite{Rajaraman} is given by $\phi_{\bar{K}}(z)=-\phi_{K}(z)$, and a
time-dependent Lorentz boosted domain kink with velocity $u$ \cite{Rajaraman}
is given by the solution%

\begin{equation}
\label{fifteen}\phi_{K}(z;u)=\phi_{0}\tanh\left[  \frac{\left(  z-z_{0}%
\right)  -ut}{w\left(  1-u^{2}\right)  ^{\frac12}}\right]  ,
\end{equation}

\noindent The spatial gradient of a static kink or antikink solution $\phi$ is%

\begin{equation}
\partial_{z}\phi=\pm\sqrt{\frac{g}{2}}\left(  \phi^{2}-\phi_{0}^{2}\right)
=\pm\sqrt{\frac{g}{2}}\phi_{0}^{2}\ \text{sech}^{2}\left(  \frac{z-z_{0}}%
{w}\right)  \label{sixteen}%
\end{equation}

\noindent which rapidly vanishes for $\left\vert z-z_{0}\right\vert \gg w$,
with $\phi_{K}\rightarrow\pm\phi_{0}$ as $\left(  z-z_{0}\right)
\rightarrow\pm\infty$.

\subsection{Kink formation}

\ \ Using the potential $V(\chi,\phi)$ given in (\ref{three}), we set $\chi=0$
and define an effective potential $U(\phi)$ for the $\phi$
field\cite{Morris95} by subtracting off the constant $V(0,0)=\frac{1}%
{4}\lambda\eta^{4}$:%

\begin{equation}
U\left(  \phi\right)  =V\left(  0,\phi\right)  -V\left(  0,0\right)  =V\left(
0,\phi\right)  -\frac{1}{4}\lambda\eta^{4}=\frac{1}{4}g\phi^{2}\left(
\phi^{2}-2\phi_{0}^{2}\right)  , \label{seventeen}%
\end{equation}

This is then used to define an effective Lagrangian for the $\phi$ field,%
\begin{equation}
L^{\left(  \phi\right)  }=\frac{1}{2}\partial^{\mu}\phi\partial_{\mu}%
\phi-U\left(  \phi\right)  . \label{19}%
\end{equation}

which, in turn, can be used to define \noindent the energy-momentum tensor
associated with the field $\phi$,%

\begin{equation}
\label{eighteen}T_{\mu\nu}^{\left(  \phi\right)  }=\partial_{\mu}\phi
\partial_{\nu}\phi-g_{\mu\nu}L^{\left(  \phi\right)  },
\end{equation}

\noindent\noindent(We can note here that the equation of motion
(\ref{thirteen}) for the field $\phi\left(  z,t\right)  $ in the string core
is obtained from the variation of $L^{\left(  \phi\right)  }$.) Therefore the
energy density of the $\phi$ field in the string core is%

\begin{equation}
\label{twenty}T_{00}^{\left(  \phi\right)  }=\frac12\left[  \left(
\partial_{0}\phi\right)  ^{2}+\left(  \partial_{z}\phi\right)  ^{2}\right]
+U\left(  \phi\right)  .
\end{equation}

\noindent For a constant field configuration $\phi=\pm\phi_{0}$ we have that
$T_{00}^{(\phi)}$ is negative, i.e., $T_{00}^{\left(  \phi\right)  }\left(
\pm\phi_{0}\right)  =U\left(  \phi_{0}\right)  =-\frac{1}{4}g\phi_{0}^{4}$,
but the total energy density in the string core (for $R\approx0$, $A_{\mu
}\approx0$, $\phi=\pm\phi_{0}$) is approximately $T_{00}=T_{00}^{\left(
\phi\right)  }\left(  \pm\phi_{0}\right)  +\frac{1}{4}\lambda\eta^{4}=\frac
{1}{4}\left(  \lambda\eta^{4}-g\phi_{0}^{4}\right)  $ which is positive,
provided that $\lambda\eta^{4}-g\phi_{0}^{4}>0$, which is assumed to be the case.

\bigskip

\ \ Inside the string core, the lowest energy configuration for the $\phi$
field is $\phi=\pm\phi_{0}$, and the $\phi$ field will tend to settle into one
of these states where either $\phi=+\phi_{0}$ or $\phi=-\phi_{0}$, but the
domains of these different states will be uncorrelated beyond some coherence
length $\xi$, which is expected to be at least as big as the width $w$ of a
kink. Two different domains must be separated by a region where $\phi=0$,
locating the center of a kink or antikink. We therefore take the average
adjacent K-\={K} separation distance to be $\xi\gtrsim w=\frac{1}{\phi_{0}%
}\left(  \frac{2}{g}\right)  ^{\frac{1}{2}}\sim1/\left(  \sqrt{g}\phi
_{0}\right)  $, and the solution $\phi_{K}$ (or $\phi_{\bar{K}}$) represents a
kink (or antikink) which separates $\pm\phi_{0}$ domains. Two consecutive
kinks are separated by an antikink, and vice versa, with an initial K-\={K}
separation distance of $\sim\xi$.

\subsection{Pressurized string}

\ \ \ \ The K and \={K} states are not expected to be static, but rather to
move around, as described by (\ref{fifteen}), and consequently begin
annihilating soon after formation. The K-\={K} annihilations give rise to a
gas of bosonic $\varphi$ particles in the string core. The $\varphi$ particles
are excitations of the $\phi$ field above the ground state, i.e., $\phi
=\phi_{0}+\varphi$, where the ground state condensate is $\phi_{0}$, say. We
denote the mass of the $\varphi$ particles inside the core by $m_{in}$. The
$\varphi$ particle mass outside the core, in vacuum, is $m_{out}=m$ [see
(\ref{three})]. For a large mass contrast where $m_{in}\ll m_{out}=m$, the
$\varphi$ bosons are largely confined to the string core, although those
particles with energies $E\gtrsim E_{esc}=m$ can eventually escape the core
and reside in the vacuum. However, particles with energy $E<m$ are trapped
inside the core. We assume here that there is indeed a high contrast in
$\varphi$ particle masses inside and outside the string, such that $m_{in}\ll
m_{out}=m,$ and the entrapped bosons form a gas with temperature $T$ inside
the string with a thermal energy distribution with $m/T=$ $\beta m\gg1$, so
that according to Bose-Einstein statistics, where the number of particles of
energy $E$ is $N(E)=\left(  e^{\beta E}-1\right)  ^{-1}$, the rate at which
$\varphi$ particles with $\beta E\gtrsim\beta m\gg1$ escape is small. The
bosonic gas exerts a pressure in the core which tends to counteract the
effects of the string tension. We refer to the string whose core is filled
with the $\varphi$ boson gas as a \textit{pressurized} string. A closed loop
of pressurized string can form through intercommutation processes or
self-intersecting string trajectories. The pressurized string loop can be
roughly thought of as resembling a (not too leaky) tire inner tube pressurized
with air. Furthermore, we consider the case where the $\varphi$ gas is assumed
to be relativistic, and can therefore have nonnegligible energy density $\rho
$, number density $n$, and pressure $p$.\textbf{\ }

\section{Loop stabilization}

\ \ We contemplate a scenario where a loop of cosmic string forms with many
K-\={K} states in the core, which begin annihilating to produce a $\varphi$
boson gas within the string, with $m_{in}\ll m_{out}=m$, so that the bosons
are effectively trapped within the string core and exert a pressure $p$. When
the effect of the pressure, which tends to expand the loop, becomes comparable
to that of the string tension, which tends to shrink the loop, an equilibrium
can be reached at some loop length $L=2\pi\mathcal{R}$ for a circular loop.
During a shrinking process, energy is radiated away in the form of $\chi$
bosons and gravitational radiation. The equilibrium configuration minimizes
the configuration energy $\mathcal{E}$, since the radial force on the loop is
$F_{R}=-\delta\mathcal{E}/\delta\mathcal{R}=0$. (We do not consider a loop
with a nonzero angular momentum of the boson gas.)

\bigskip

\ \ The energy of the loop near equilibrium is the sum of two terms,%
\begin{equation}
\mathcal{E}=\mu L+\rho_{G}A_{s}L \label{23}%
\end{equation}

where $\mu$ is the energy per unit length (tension) of the string (excluding
the boson gas), $\rho_{G}$ is the energy density of the $\varphi$ gas,
$A_{s}=\pi r_{0}^{2}$ is the cross sectional area of the string, and
$L=2\pi\mathcal{R}$ is the length of the circular loop with radius
$\mathcal{R}$. We take the $\varphi$ bosons to be effectively massless, with
the loop able to equilibrate at a temperature $m_{in}\ll T\ll m$. As a loop
physically shrinks, energy is lost and the radial force $F_{R}<0$ until a
physical equilibrium is reached. At that time, the loop energy is given by
(\ref{23}), with $L=$ const. This equilibrium value of $L$ is determined by
minimizing $\mathcal{E}$ with respect to $L$. Care must be taken in this
variation, however. We consider a \textit{virtual} variation of the loop
length $L$ while holding entropy $S$ and $\varphi$ boson particle number $N$
fixed. (One implies the other, since both are proportional to $T^{3}A_{s}L$.)
Using $\rho_{G}=\frac{\pi^{2}}{30}T^{4}$ for the relativistic $\varphi$ boson
gas, (\ref{23}) becomes%
\begin{equation}
\mathcal{E}=\mu L+\left(  \frac{\pi^{2}}{30}T^{4}\right)  A_{s}L \label{24}%
\end{equation}

The entropy density of the gas is\cite{KTbook} $s=\frac{2\pi^{2}}{45}T^{3}$
and the number density is\cite{KTbook} $n=\frac{\zeta(3)}{\pi^{2}}T^{3}$, so
that holding entropy $S=sA_{s}L$ and particle number $N=nA_{s}L$ constant
during the virtual variation of $L$ results in the equilibrium constraint
$T^{3}L=$ const:%
\begin{equation}
\frac{N}{A_{s}}=\frac{\zeta(3)}{\pi^{2}}T^{3}L=c_{1};\ \ \ \ \ \frac{S}{A_{s}%
}=\frac{2\pi^{2}}{45}T^{3}L=c_{2};\ \ \ \ T^{3}L=\frac{N\pi^{2}}{\zeta
(3)A_{s}}=\frac{45S}{2\pi^{2}A_{s}}\equiv C \label{25}%
\end{equation}

Therefore, at equilibrium, the entropy per $\varphi$ particle is%
\begin{equation}
\frac{S}{N}=\frac{s}{n}=\frac{c_{2}}{c_{1}}=\frac{2\pi^{4}}{45\zeta(3)}%
\approx3.6 \label{26}%
\end{equation}

and from (\ref{25})%
\begin{equation}
T=C^{1/3}L^{-1/3} \label{27}%
\end{equation}

Using (\ref{27}) in (\ref{24}) gives an equilibrium loop energy%
\begin{align}
\mathcal{E}  &  =\mu L+\frac{\pi^{2}}{30}A_{s}(T^{3}L)T=\mu L+\frac{\pi^{2}%
}{30}CA_{s}(C^{1/3}L^{-1/3})\nonumber\\
&  =\mu L+\frac{\pi^{2}A_{s}C^{4/3}}{30}L^{-1/3}=\mu L+C_{1}L^{-1/3}
\label{28}%
\end{align}

where $C_{1}=\pi^{2}A_{s}C^{4/3}/30$. Minimizing this with respect to $L$
results in%
\begin{equation}
\frac{\partial\mathcal{E}}{\partial L}=\mu-\frac{1}{3}C_{1}L^{-4/3}=0\implies
L=\left(  \frac{C_{1}}{3\mu}\right)  ^{3/4},\ \ \ L^{1/3}=\left(  \frac{C_{1}%
}{3\mu}\right)  ^{1/4} \label{29}%
\end{equation}

The equilibrium configuration energy is then%
\begin{equation}
\mathcal{E}=\mu\left(  \frac{C_{1}}{3\mu}\right)  ^{3/4}+C_{1}\left(
\frac{3\mu}{C_{1}}\right)  ^{1/4}=(C_{1}^{3}\mu)^{1/4}\left[  \left(  \frac
{1}{3}\right)  ^{3/4}+3^{1/4}\right]  \approx1.75(C_{1}^{3}\mu)^{1/4}
\label{30}%
\end{equation}

The gas energy to string energy ratio is%
\begin{equation}
\frac{\mathcal{E}_{G}}{\mathcal{E}_{s}}=\frac{3^{1/4}}{\left(  \frac{1}%
{3}\right)  ^{3/4}}=\allowbreak3 \label{31}%
\end{equation}

\bigskip

\ \ \ \ \ Therefore, the energy of the equilibrated loop is%
\begin{equation}
\mathcal{E}=\mathcal{E}_{s}+\mathcal{E}_{G}=4\mathcal{E}_{s}=\frac{4}%
{3}\mathcal{E}_{G}=4\mu L=4\mu^{1/4}\left(  \frac{1}{3}C_{1}\right)
^{3/4}\propto\mu^{1/4}N \label{32}%
\end{equation}

From (\ref{25}) the loop size is given by $L=C/T^{3}$, and since
$\mathcal{E}=4\mu L=8\pi\mu\mathcal{R}$, the loop radius is%
\begin{equation}
\mathcal{R}=\frac{C}{2\pi T^{3}}=\frac{\mathcal{E}}{8\pi\mu} \label{33}%
\end{equation}

The Schwarzschild radius is $R_{S}=2G\mathcal{E}=2\mathcal{E}/M_{P}^{2}$, so%
\begin{equation}
\frac{\mathcal{R}}{\mathcal{R}_{S}}=\frac{1}{16\pi G\mu};\ \ \ \ \frac
{\mathcal{R}}{\mathcal{R}_{S}}>1\implies G\mu<\frac{1}{16\pi} \label{34}%
\end{equation}

The condition that the equilibrated loop radius lies outside the Schwarzschild
radius manifests itself as an upper bound on the string tension, $G\mu
<1/16\pi$. This upper bound lies well above present observational constraints
for the string tension, with $G\mu\lesssim6.4\times10^{-7}$ for Abelian-Higgs
strings\cite{Battye}, and therefore does not exclude any of the observational
parameter space for the string tension.

\bigskip

\ \ Now suppose that the loop is created with an initial radius less than the
equilibrium radius, $\mathcal{R}_{0}<\mathcal{R}$. What happens in this case?
Again, consider a \textit{virtual} change in radius $\mathcal{R}_{0}$ while
holding $N_{0}$, $S_{0}$ fixed, so that again we have the
\textit{instantaneous} constraint in (\ref{27}) for the virtual deformation.
Then by (\ref{28}) we have%
\begin{equation}
\mathcal{E}_{0}=\mu L_{0}+C_{1}L_{0}^{-1/3} \label{36}%
\end{equation}

Now the radial components of the string tension and gas pressure forces,
$F_{R,0}^{s}$ and $F_{R,0}^{G}$, respectively, are%
\begin{equation}
F_{R,0}^{s}=-\frac{\partial\mathcal{E}_{0,s}}{\partial\mathcal{R}_{0}}%
=-2\pi\mu;\ \ \ \ \ F_{R,0}^{G}=-\frac{\partial\mathcal{E}_{0,G}}%
{\partial\mathcal{R}_{0}}=+\frac{1}{3}(2\pi)^{-1/3}C_{1}\mathcal{R}_{0}^{-4/3}
\label{37}%
\end{equation}

If $|F_{R,0}^{G}/F_{R,0}^{s}|>1$, the loop will expand, assuming, of course,
$\mathcal{R}_{0}>\mathcal{R}_{S}$, and gas energy is converted into string
energy in the process until $|F_{R}^{G}/F_{R}^{s}|=1$ at equilibrium. If
$|F_{R,0}^{G}/F_{R,0}^{s}|<1$, the loop shrinks, emits $\chi$ boson and
gravitational radiation until it falls through its Schwarzschild radius,
ending up as a black hole.

\bigskip

\ \ We see from (\ref{28}) and (\ref{29}) that the equilibrium length $L$
determined there corresponds to a point of stable equilibrium, so that for a
dynamical loop with time varying radius, the loop radius will oscillate about
the static eqilibrium value (assuming no loss of $\varphi$ bosons), implying
short term stability. However, in the longer term loop energy will be lost due
to a leakage of high energy $\varphi$ bosons, along with $\chi$ boson and
gravitational radiation, leading to eventual loop decay. As long as the
temperature is low enough, we might expect the loop to lose $\varphi$ bosons
at only a slow rate, and the loop may stay fairly close to equilibrium
(although $\chi$ boson and gravitational radiation are expected due to loop
oscillations). We may argue that the $\varphi$ boson gas prefers to remain
inside the string core in the form of $N$ $\varphi$ bosons rather than outside
the loop in the form of $N$ $\phi$ bosons. This is because (provided the
temperature $T$ is not too high) the pressurized string loop will have a
configuration energy $\mathcal{E}$ which will be less than that of an
unpressurized string loop with mass $\mathcal{E}_{s}$, plus $N$ $\phi$ bosons
(each of mass $m$ in vacuum outside the loop) with a minimal total energy of
$\mathcal{E}_{s}+Nm$. Specifically, we require $\mathcal{E}_{G}<Nm$ for
stability against a complete, rapid evaporation of the $\varphi$ gas. This
translates into the condition%
\begin{equation}
\frac{\mathcal{E}_{G}}{Nm}=\frac{B}{D}\frac{NT}{Nm}=\frac{B}{D}\frac{T}%
{m}<1;\ \ T<\frac{D}{B}m=.37m \label{Tm}%
\end{equation}

where $N=\frac{\zeta(3)}{\pi^{2}}T^{3}A_{s}L\equiv DT^{3}L$ and $\mathcal{E}%
_{G}=\left(  \frac{\pi^{2}}{30}T^{4}\right)  A_{s}L\equiv BT^{4}L$ and we
define%
\begin{equation}
B=\frac{\pi^{2}A_{s}}{30},\ \ \ D=\frac{\zeta(3)A_{s}}{\pi^{2}},\ \ \frac
{D}{B}=\frac{30\zeta(3)}{\pi^{4}}=.37 \label{BD}%
\end{equation}

Therefore, at low temperatures $T\ll m$ there may be a slow leakage of
$\varphi$ bosons, but at higher temperatures we expect a departure from
equilibrium, with a higher leakage rate.

\section{Loop decay}

\ \ Even with a stabilization mechanism as described above, we expect a loop
to become unstable over longer periods of time. This is because a string loop
can lose energy through the escape of high energy $\varphi$ particles of mass
$m_{in}$ into the vacuum outside the string, where they become $\phi$
particles with mass $m$. An oscillating loop will also lose energy due to
gravitational radiation\cite{Allen} at a rate $P_{\text{rad}}\sim\gamma
G\mu^{2}$, where $\gamma$ depends on the string trajectory, with $\gamma
\sim100$ being typical. The oscillating loop may also lose energy through
$\chi$ boson radiation, but the $\chi$ boson radiation must come from
nontraveling waves in the string\cite{VEV}. (A string can, however, support
certain traveling waves of arbitrary amplitude with no $\chi$ boson
radiation\cite{VV}.)

\bigskip

\ \ Let us focus here on the process of loop decay through $\phi$ boson
radiation. For a simplified and qualitative approach, it is assumed that a
$\varphi$ particle in the string gas with energy $E\geq m$ can exit the string
and appear as a $\phi$ particle in the vacuum. A loop initially near
equilibrium with mass $\mathcal{E}_{0}$ and $\varphi$ number $N_{0}$ at some
initial time $t_{0}$ will eventually lose energy through $\phi$, $\chi$, and
gravitational radiation, so that at a later time $t$ the loop mass is
$\mathcal{E}$ and the $\varphi$ number is $N$. With the simple assumption that
the number $n_{\phi}$ of $\phi$ particles produced in the time interval
$t-t_{0}$ is given by $n_{\phi}=N_{0}-N$, we can write $\mathcal{E}%
_{0}=\mathcal{E}+n_{\phi}m+\mathcal{E}_{X}=\mathcal{E}+(N_{0}-N)m+\mathcal{E}%
_{X}$, where $\mathcal{E}_{X}$ represents $\phi$ kinetic energy, along with
$\chi$ and gravitational energy that is produced by the decaying loop. We
rewrite this as%
\begin{equation}
mN-\mathcal{E}=K+\mathcal{E}_{X}\equiv Q(t) \label{Q}%
\end{equation}

where the constant $K=mN_{0}-\mathcal{E}_{0}$ and the function
$Q(t)=K+\mathcal{E}_{X}(t)$ is a monotonically increasing function of time.
(We note that the constant $K$ depends upon the initial loop mass and
$\varphi$ particle number, i.e., the loop parameters near the equilibrium
state, and we therefore do not regard $K$ as a universal constant for all
loops in equilibrium. Each loop has its own value for $K$.) From (\ref{24}),
(\ref{25}), and (\ref{BD}) we have%
\begin{equation}
\mathcal{E}=\left[  \mu+\left(  \frac{\pi^{2}}{30}T^{4}\right)  A_{s}\right]
L\equiv(\mu+BT^{4})L,\ \ \ \ \ N=\frac{\zeta(3)A_{s}}{\pi^{2}}T^{3}L\equiv
DT^{3}L \label{37a}%
\end{equation}

where%
\begin{equation}
B=\frac{\pi^{2}A_{s}}{30},\ \ \ \ \ D=\frac{\zeta(3)A_{s}}{\pi^{2}%
},\ \ \ \frac{B}{D}=2.7,\ \ \ \ \frac{D}{B}=.37 \label{37b}%
\end{equation}

The condition (\ref{Q}) can now be written as%

\begin{equation}
mDT^{3}-BT^{4}=\mu+\frac{Q}{L} \label{37c}%
\end{equation}

For a temperature $T\ll m$, the first term on the left hand side of
(\ref{37c}) dominates, so that%

\begin{equation}
T\approx\left[  \frac{1}{mD}\left(  \mu+\frac{Q}{L}\right)  \right]
^{1/3},\ \ \ (T\ll m) \label{T}%
\end{equation}

As the loop evolves, it loses energy and shrinks and the loop temperature $T$
increases. (The constant $K$, and therefore $Q$, is positive. This can be seen
from (\ref{37a}) and (\ref{32}), since $\mathcal{E}_{0}\approx\frac{4}%
{3}\mathcal{E}_{G}=\frac{4}{3}BT^{4}L_{0}\ll N_{0}m$ for $T\ll m$.)

\bigskip

\ \ However, as $T$ increases (\ref{37c}) must break down beyond some value
$T_{m}$, where our simple assumptions apparently begin to fail. One can see
that this is the case by noticing that as $L$ decreases and $T$ increases, the
right hand side (RHS) of (\ref{37c}) is monotonically increasing, but the left
hand side (LHS) begins to decrease for $T>T_{m}$. The temperature $T_{m}$
therefore locates the local maximum of the LHS, which is found to occur at
$T_{m}=\frac{3D}{4B}m\approx\frac{1}{4}m$. \ Therefore, the temperature $T$
and loop length $L$ are approximately related by (\ref{37c}) for $T\lesssim
T_{m}\approx\frac{1}{4}m$, where we expect a more rapid rate of production of
$\phi$ bosons to begin occurring. This value of $T_{m}$ is close to the
temperature found in (\ref{Tm}) where the loop begins to lose its bosonic stabilization.

\bigskip

For the loop near equilibrium with mass $\mathcal{E}_{0}$ and temperature
$T_{0}$, we have $\mathcal{E}_{0}=4\mu L_{0}=\frac{4}{3}BT_{0}^{4}L_{0}$ which
gives%
\begin{equation}
T_{0}=\left(  \frac{3\mu}{B}\right)  ^{1/4}=\left(  \frac{90\mu}{\pi^{3}%
r_{0}^{2}}\right)  ^{1/4}\sim\left(  \mu m_{\chi}^{2}\right)  ^{1/4}=\left(
G\mu M_{P}^{2}m_{\chi}^{2}\right)  ^{1/4}\lesssim3\times10^{-2}\left(
M_{P}m_{\chi}\right)  ^{1/2} \label{Tzero}%
\end{equation}

where the Planck mass is $M_{P}=1/\sqrt{G}$ and $G\mu\lesssim6.7\times10^{-7}$
has been used\cite{Battye}. This temperature is independent of loop size.
Therefore all loops approach an equilibration at this temperature.

\bigskip

\ \ We then form the following qualitative picture of loop decay. A metastable
loop begins to slowly shrink, and as it shrinks the temperature rises in
accordance with (\ref{T}), as long as $T\ll m$. As the temperature rises, the
rate of $\phi$ boson emission increases and the rate of decay increases.
However, our approximations and assumptions leading to (\ref{T}) are expected
to break down as $T$ approaches $m$, and the loop is expected to depart from
being near equilibrium with a rapid release of $\phi$ radiation, along with
$\chi$ particle and gravitational radiation.

\bigskip

\ \ We would like to obtain a rough estimate of how long the bosonic
stabilization mechanism might persist. In order to do so, we treat the loop as
a $\phi$ blackbody that emits $\varphi$ particles having energy $E\geq m$. The
$\varphi$ particles are taken to be effectively massless. The energy density
of $\varphi$ particles above this energy threshold is \cite{Reif}%
\begin{equation}
u(T)=\frac{T^{4}}{2\pi^{2}}\int_{\beta m}^{\infty}\frac{\eta^{3}}{e^{\eta}%
-1}d\eta=\frac{I(\beta m)}{2\pi^{2}}T^{4} \label{u1}%
\end{equation}

where $\eta=\beta E$. (This is 1/2 the corresponding photon energy density,
since photons have two polarization states.) For temperatures $T\ll m$, i.e.,
$\beta m\gg1$,%
\begin{equation}
I(\beta m)\approx\int_{\beta m}^{\infty}\eta^{3}e^{-\eta}d\eta\approx\left(
\beta m\right)  ^{3}e^{-\beta m}\ll1,\ \ \ \ \ \ \ (\beta m\gg1) \label{I}%
\end{equation}

On the other hand, for a temperature $T_{\ast}=.37m$, i.e., $\beta_{\ast
}m=2.7$, as given in (\ref{Tm}), we have%
\begin{equation}
I(2.7)=\int_{2.7}^{\infty}\frac{\eta^{3}}{e^{\eta}-1}d\eta\approx4.4
\label{I2}%
\end{equation}

Therefore, for the $\varphi$ boson gas temperatures of interest we take,
roughly, $I(\beta m)\lesssim1$. The radiated power of $\phi$ bosons per unit
area of the loop is \cite{Reif}%
\begin{equation}
\mathcal{P}_{\phi}(T)=\frac{1}{4}u(T)=\frac{1}{A_{\text{loop}}}\frac
{d\mathcal{E}}{dt}\lesssim\frac{T^{4}}{8\pi^{2}} \label{P}%
\end{equation}

To obtain a very rough estimate for the lifetime of the bosonic stabilization
mechanism to work, we set $\mathcal{E}\sim\mathcal{E}_{0}\approx\frac{4}%
{3}BT_{0}^{4}L_{0}$, $A_{\text{loop}}\sim2\pi r_{0}L_{0}$, and $\mathcal{P}%
_{\phi}\sim d\mathcal{E}/dt\sim\mathcal{E}_{0}/\tau$. This results in%
\begin{equation}
\tau\sim\frac{\mathcal{E}_{0}}{\mathcal{P}_{\phi}A_{\text{loop}}}\sim
\frac{2BT_{0}^{4}}{3\pi r_{0}}\cdot\frac{8\pi^{2}}{T_{0}^{4}}\sim\frac{16\pi
B}{3r_{0}}\sim\frac{16\pi}{3r_{0}}\cdot\frac{\pi^{3}r_{0}^{2}}{30}\sim
\frac{8\pi^{4}}{45}m_{\chi}^{-1} \label{tau}%
\end{equation}

where $r_{0}\sim m_{\chi}^{-1}$ has been used along with the definition of $B$
in (\ref{BD}). This lifetime is independent of the initial loop size $L_{0}$.
This lifetime would be on the order of $10^{-14}$ seconds for $m_{\chi}\sim1$
eV, on the order of $10^{-23}$ seconds for $m_{\chi}\sim1$ GeV, and even
smaller for a more massive $\chi$ boson.

\bigskip

\ \ An ordinary oscillating loop loses energy through gravitational
radiation\cite{Allen} at a rate $P_{\text{rad}}\sim\gamma G\mu^{2}$. Setting
this equal to $\mathcal{E}_{0}/\Delta t\sim4\mu L_{0}/\Delta t$ gives $\Delta
t\sim4L_{0}/(\gamma G\mu)$ as the lifetime of a loop due to gravitational
energy loss. A comparison of $\tau$ and $\Delta t$ gives%
\begin{equation}
\frac{\tau}{\Delta t}\sim\frac{2\pi^{4}}{45}\frac{\gamma G\mu}{m_{\chi}L_{0}%
}\sim\frac{2\pi^{4}(\gamma G\mu)}{45}\frac{r_{0}}{L_{0}}\sim4\gamma G\mu
\frac{r_{0}}{L_{0}}\lesssim10^{-4}\frac{r_{0}}{L_{0}} \label{ratio}%
\end{equation}

where $r_{0}/L_{0}<1$ and $G\mu\lesssim6.4\times10^{-7}$ has been
used\cite{Battye}, along with $\gamma\sim10^{2}$. It therefore appears that
the bosonic stabilizing mechanism is effective for much less than one ten
thousandth of the lifetime of the loop. The loop evidently loses its $\varphi$
boson gas long before decaying away.

\bigskip

\ \ We can notice that when we set the lower limit of the integral in
(\ref{u1}) to zero, we obtain $I(0)=\pi^{4}/15\approx\allowbreak
6.\,\allowbreak5$, and in this case $u(T)=\rho_{G}$. An examination of
(\ref{I}) and (\ref{I2}) shows that initially, for the loop near the
equilibrium state, we have $u/\rho_{G}\ll1$, but at $T\sim T_{\ast}=.37m$ we
have $u/\rho_{G}\sim2/3$. So initially only a small fraction of the $\varphi$
gas can escape the string core, but the temperature rapidly increases and at
$T\sim T_{\ast}$ most of the gas can escape.

\section{Particle Masses and model parameters}

\ \ Let us calculate the $\varphi$ boson mass $m_{in}$ and compare it to the
kink mass $M_{K}$. From (\ref{seventeen}) and (\ref{19}), with $R=0$ in the
string core, and with $\phi=\phi_{0}+\varphi$,%
\begin{equation}
U\left(  \phi\right)  =\frac{1}{4}g\phi^{2}\left(  \phi^{2}-2\phi_{0}%
^{2}\right)  =\frac{1}{4}g\left(  \phi^{4}-2\phi_{0}^{2}\phi^{2}\right)
\label{38}%
\end{equation}

Then $m_{in}^{2}=\partial^{2}U/\partial\varphi^{2}|_{\varphi=0}=\partial
^{2}U/\partial\phi^{2}|_{\phi=\phi_{0}}=2g\phi_{0}^{2}$. Using (\ref{eight})
for the definition of $\phi_{0}$ gives%

\begin{equation}
m_{in}=\sqrt{2g}\phi_{0}=\sqrt{2(f\eta^{2}-m^{2})} \label{40}%
\end{equation}

\ \ For a large mass contrast $m_{in}/m\ll1$, we require%
\begin{equation}
2(f\frac{\eta^{2}}{m^{2}}-1)\ll1,\ \ \text{i.e.,}\ \ \ 2g\frac{\phi_{0}^{2}%
}{m^{2}}\ll1 \label{41}%
\end{equation}

i.e., $\sqrt{g}\phi_{0}\ll m$, which can be satisfied for $m^{2}\sim
O(f\eta^{2})$, with $m^{2}<f\eta^{2}$.

\bigskip

\ \ The mass of a kink is\cite{Morris95}%
\begin{equation}
M_{K}\sim\frac{\sqrt{g}\phi_{0}^{3}}{\lambda\eta^{2}} \label{Kmass}%
\end{equation}

This is obtained from the energy density $T_{00}^{(K)}$ of a static kink,
which in turn is obtained from the energy density of the $\phi$ field in
excess of that due to the condensate $\phi=\phi_{0}$ for a string containing
only the condensate and kinks and antikinks:%
\begin{equation}
T_{00}^{(K)}\equiv T_{00}^{(\phi)}(\phi_{K})-T_{00}^{(\phi)}(\pm\phi
_{0})=(\partial_{z}\phi_{K})^{2}=\frac{1}{2}g\phi_{0}^{4}\text{sech}%
^{4}\left(  \frac{z-z_{0}}{w}\right)  \label{Kenergy}%
\end{equation}

(This corrects a typo in Eq.(22) of Ref.\cite{Morris95}.) An integration of
$T_{00}^{(K)}$ over $z$ gives the kink surface energy density $\Sigma
_{K}=\frac{4}{3}\sqrt{\frac{g}{2}}\phi_{0}^{3}$. (This corrects Eq.(23) of
Ref.\cite{Morris95}.) The kink mass is then $M_{K}\approx\Sigma_{K}r_{0}^{2}$,
where $r_{0}\approx m_{\chi}^{-1}=\left(  2\lambda\eta^{2}\right)  ^{-1/2}$ is
the radius of the string itself. The result is (\ref{Kmass}).

\bigskip

\ \ Finally, for the K, \={K} particles to be much heavier than the $\varphi$
particles in the string, so that \={K}K annihilations producing $\varphi$
bosons will dominate over the reverse where \={K}K pairs are formed, we
require%
\begin{equation}
\frac{m_{in}}{M_{K}}\sim\frac{\lambda\eta^{2}}{\phi_{0}^{2}}\ll
1,\ \ \text{\ \ or\ \ \ \ }\lambda\ll\frac{\phi_{0}^{2}}{\eta^{2}}
\label{mratio}%
\end{equation}

where (\ref{40}) and (\ref{Kmass}) have been used. Since $m_{\chi}^{2}%
\sim\lambda\eta^{2}$, the above result can also be written as $\phi_{0}^{2}\gg
m_{\chi}^{2}$. If (\ref{mratio}) is not obeyed, we expect the string core to
contain a mixed gas of kinks and $\varphi$ bosons. The gas is again treated as
radiation as long as both the kink and $\varphi$ masses are small compared to
the temperature $T$, i.e., $\beta M_{K}\ll1$ and $\beta m_{in}\ll1$, so the
basic mechanism for stabilization of a loop would still apply.

\bigskip

\ \ Some of the parameters and constraints for the model are summarized below.%

\begin{equation}%
\begin{array}
[c]{ll}%
m_{\chi}=\sqrt{2\lambda}\eta;\ m_{out}=m;\ \ m_{in}=\sqrt{2g}\phi
_{0};\ \smallskip\ M_{K}\sim\dfrac{\sqrt{g}\phi_{0}^{3}}{\lambda\eta^{2}} & \\
\phi_{0}=\sqrt{(f\eta^{2}-m^{2})/g};\ w\sim\dfrac{1}{\sqrt{g}\phi_{0}%
},\ \ \ \ \xi\gtrsim w\smallskip & \\
\dfrac{m_{in}}{m}\ll1\implies\sqrt{g}\phi_{0}\ll m\implies\sqrt{(f\eta
^{2}-m^{2})}\ll m\smallskip & \\
T_{00}>0\implies\sqrt{\lambda}\eta^{2}>\sqrt{g}\phi_{0}^{2}:\ \ m_{\chi}%
^{2}/\sqrt{\lambda}>m_{in}^{2}/\sqrt{g}\smallskip & \\
&
\end{array}
\label{parameters}%
\end{equation}

\section{Summary}

\ \ A model of an Abelian-Higgs gauge string has been considered where there
is an additional interaction of the scalar string field ($\chi$) with a
real-valued scalar field ($\phi$). There exists a range of parameters for the
model for which a $\phi$ condensate will tend to form in the string's core,
settling into the low energy states $\phi=\pm\phi_{0}$, breaking a $Z_{2}$
symmetry. However, $+\phi_{0}$ and $-\phi_{0}$ domains will be uncorrelated
beyond some coherence length $\xi$, and there will be regions where $\phi=0$
in between different domains. These $\phi=0$ regions locate kinks and
antikinks that reside within the string core.

\bigskip

\ \ The kinks and antikinks will collide and undergo annihilations, producing
$\varphi$ bosons, which are excitations of the $\phi$ field within the string
core, $\phi=\phi_{0}+\varphi,$ for instance. The mass $m_{in}$ of the
$\varphi$ particles inside the string is not the same as the mass $m$ of the
$\phi$ particles outside the string (where the vacuum state is located by
$\phi=0$). The case considered here is that for which $m_{in}\ll m$ and
$m_{in}\ll M_{K}$, where $M_{K}$ is the kink mass. In this case, $\varphi$
bosons are produced from K-\={K} annihilations, and get trapped within the
string core, giving rise to a gas pressure and additional energy. The gas
pressure tends to cause the expansion of a closed string loop, while the
string tension tends to cause a contraction. An equilibrium is established
when these two effects are balanced and the configuration energy is minimized,
provided that no $\varphi$ particles exit the string core. The equilibrium
loop radius $\mathcal{R}$ is found to be given by the ratio of the string mass
$\mathcal{E}$ and the string tension $\mu$, i.e., $\mathcal{R}=\mathcal{E}%
/(8\pi\mu)=C/(2\pi T^{3})$ [see Eq.(\ref{33})], where \ $C$ is a constant
depending on the number $N$ of $\varphi$ particles in the string loop.

\bigskip

\ \ However, high energy $\varphi$ bosons with energies $E\gtrsim m$ will
escape the string core, so that the string loop is unstable over longer
periods of time. The loop begins to shrink, relatively slowly at first, but at
a rapidly increasing rate. An oscillating loop will also lose energy due to
gravitational radiation\cite{Allen} at a rate $P_{\text{rad}}\sim\gamma
G\mu^{2}$, where $\gamma$ depends on the string trajectory, with $\gamma
\sim100$ being typical. The oscillating loop can also lose energy through
$\chi$ boson radiation due to loop oscillations, but the $\chi$ boson
radiation must come from nontraveling waves (e.g., standing waves) in the
string\cite{VEV}. (A string can, however, support certain traveling waves of
arbitrary amplitude with no $\chi$ boson radiation\cite{VV}.) However,
estimates based upon the rate of $\phi$ particle emission by the loop indicate
that the bosonic stabilization mechanism considered here is rather
ineffective, lasting for a short duration which is much less than the lifetime
of the loop due to gravitational radiation.

\bigskip

\textbf{Acknowledgment: }I thank an anonymous referee for valuable comments.

\end{document}